\documentclass[conference]{IEEEtran}
\usepackage{amsmath}

\usepackage{color}
\usepackage{framed}
\usepackage{graphicx}
\usepackage{amsfonts}
\usepackage{epstopdf}
\usepackage[caption=false]{subfig}

\usepackage{enumitem}
\usepackage{cite}

\usepackage{hyperref}
\definecolor{MyDarkBlue}{rgb}{0,0.08,0.50}
\definecolor{BrickRed}{rgb}{0.65,0.08,0}
\definecolor{Black}{rgb}{0,0,0}
\hypersetup{
   colorlinks=true,       % false: boxed links; true: colored links
   linkcolor=Black,          % color of internal links
   citecolor=Black,        % color of links to bibliography
   filecolor=Black,      % color of file links         % color of external links
   urlcolor=Black
}
\usepackage{url}

\ifCLASSINFOpdf
\else
\fi
\IEEEoverridecommandlockouts
\IEEEpubid{\makebox[\columnwidth]{Preprint submitted to arXiv  \hfill} \hspace{\columnsep}\makebox[\columnwidth]{ }}

\begin{document}
\title{Adaptive Broadcast Suppression\\ for Trickle-Based Protocols}

\author{\IEEEauthorblockN{Thomas M.M. Meyfroyt, Milosh Stolikj, Johan J. Lukkien}
\IEEEauthorblockA{Department of Mathematics \& Computer Science\\
Eindhoven University of Technology\\
P.O. Box 513, 5600 MB Eindhoven, The Netherlands\\
Email: \{t.m.m.meyfroyt, m.stolikj, j.j.lukkien\}@tue.nl}
}

\maketitle

\begin{abstract}
Low-power wireless networks play an important role in the Internet of Things. Typically, these networks consist of a very large number of lossy and low-capacity devices, challenging the current state of the art in protocol design. In this context the Trickle algorithm plays an important role, serving as the basic mechanism for message dissemination in notable protocols such as RPL and MPL. While Trickle's broadcast suppression mechanism has been proven to be efficient, recent work has shown that it is intrinsically unfair in terms of load distribution and that its performance relies strongly on network topology. This can lead to increased end-to-end delays (MPL), or creation of sub-optimal routes (RPL). Furthermore, as highlighted in this work, there is no clear consensus within the research community about what the proper parameter settings of the suppression mechanism should be. We propose an extension to the Trickle algorithm, called \textit{adaptive-}$\boldsymbol{k}$, which allows nodes to individually adapt their suppression mechanism to local node density. Supported by analysis and a case study with RPL, we show that this extension allows for an easier configuration of Trickle, making it more robust to network topology.
\end{abstract}
%\begin{keywords}performance evaluation, protocol design, RPL, Trickle, wireless sensor networks \end{keywords}

\section{Introduction}
In the novel Internet of Things (IoT) paradigm, daily objects called smart objects will form large meshed wireless networks that are able to gather and exchange large amounts of data. The size of these networks makes their deployment an undertaking which challenges the current state of the art in network design and networking protocols.

A specific flooding algorithm, called Trickle \cite{Levis2004}, has been introduced for its suitability in this context. The Trickle algorithm is based on stochastic timers and a window doubling mechanism to balance traffic load and response times. While Trickle was originally designed for propagating and maintaining code updates in wireless sensor networks, it has shown to be a powerful mechanism that can be applied to a wide range of protocol design problems and therefore has been documented by the Internet Engineering Task Force (IETF) in its own RFC \cite{Trickle-RFC}.  From then, it has been adopted in many other protocols. Notable protocols that use Trickle are the IPv6 Routing Protocol for Low Power and Lossy networks (RPL) \cite{RPL-RFC} and the IPv6 Multicast protocol for Low Power and Lossy Networks (MPL) \cite{MPL-draft}. In RPL, Trickle is used to control the transmission of routing control information. MPL uses Trickle to forward multicast packets in constrained networks and is currently being standardized.

Because of the algorithm's novelty, optimizing its usage and design are important ongoing topics and the focus of many research efforts. Many aspects of Trickle's performance are already well understood, however some outstanding issues remain. One of these issues is that little is known about the influence of Trickle's broadcast suppression mechanism on performance measures such as end-to-end delay and load distribution, as has been pointed out by recent works \cite{Kermajani2014-2, La2014}. Consequently, the current recommendations on how to configure Trickle's suppression mechanism are very conservative.

The main contribution of this paper is threefold. Firstly, we shed light on why Trickle's suppression mechanism is badly understood and why it is hard to provide recommendations for its correct configuration. Secondly, we propose an extension to Trickle, called \textit{adaptive-}$k$, that helps overcome these issues. This extension allows nodes to adapt their suppression mechanism according to local density information. Thirdly, both analytically and by simulations, we show that this extension has several advantages compared to the original Trickle algorithm: i) it leads to a more fair load distribution among nodes; ii) it ensures good functionality of the Trickle suppression mechanism; iii) it makes Trickle capable of adapting its suppression mechanism to network topologies of varying densities. As a case study, we implement \textit{adaptive-}$k$ Trickle as part of the RPL protocol. Simulation and physical experiments show that these improvements lead to easier configuration and better performance, allowing RPL to easily discover good routes, while suppressing many redundant control messages.

This paper is organized as follows: in Section \ref{trickle} we provide a description of the Trickle algorithm. We then give a detailed overview of related work in Section \ref{relatedwork} and discuss which parts of the Trickle algorithm are not well studied. In Section \ref{adaptive} we take a closer look at the Trickle suppression mechanism and discuss why setting it correctly is difficult. Additionally, we propose \textit{adaptive-}$k$, an extension to the Trickle algorithm which allows nodes to adapt this parameter according to local information on node density. In Section \ref{analysis} we present analysis and simulations of special network topologies which highlight the benefits of using \textit{adaptive-}$k$. Finally, as a case study, in Section \ref{RPL} we compare the performance of RPL while using the original and the extended Trickle algorithm. We summarize our results in Section \ref{conclusion}.

\section{The Trickle Algorithm}\label{trickle}
We now describe the Trickle algorithm as in \cite{Meyfroyt2014-4}. The Trickle algorithm has three parameters:
 \begin{itemize}[noitemsep]
 \item A threshold value $k$, called the redundancy constant.
 \item The minimum interval length $I_{\min}$.
 \item The maximum interval length $I_{\max}$.
 \end{itemize}
Furthermore, each node in the network has its own timer and keeps track of three local variables:
  \begin{itemize}[noitemsep]
 \item The current interval length $I$.
 \item A counter $c$, counting the number of messages heard during an interval.
 \item A broadcasting time $t$ during the current interval.
 \end{itemize}
The behavior of each node is described by the following set of rules:
\begin{enumerate}[noitemsep]
 \item At the start of a new interval a node resets its timer and counter $c$ and sets $t$ to a value in $[I/2,I]$ uniformly at random.
 \item When a node hears a message that is consistent with the information it has, it increments $c$ by 1.
 \item When a node's timer reaches time $t$, the node broadcasts its message if $c<k$.
 \item When a node's timer reaches time $I$, it increases its interval length to $\min(2I, I_{\max})$ and starts a new interval.
 \item When a node hears a message that is inconsistent with its own information, then if $I>I_{\min}$ it sets $I$ to $I_{\min}$ and starts a new interval, otherwise it does nothing.
\end{enumerate}
What constitutes a consistent or an inconsistent message, can be determined by the user and depends on the setting and/or application in which the algorithm is used.

Even though the Trickle algorithm is easily understood and implemented, the configuration of the three parameters $k$, $I_{\min}$ and $I_{\max}$ is left to the user. This raises the question of how one can optimally configure these parameters. Of course, this depends on many things, such as network topology, the application, link-layer characteristics etc., which makes giving good recommendations difficult.

In this work we focus on evaluating how one should configure the redundancy constant $k$. In the following section we first give a detailed overview of the related work carried out in this regard.

\section{Related Work}\label{relatedwork}
Since Trickle has become such an important protocol for the IoT, much research has been devoted to analyzing its performance and optimizing its usage and design. We list some of the related works that consider Trickle, both when used for flooding and when used in RPL.

\subsection{Trickle as a flooding mechanism}
The strength of Trickle's polite gossip policy that suppresses redundant transmissions depending on the redundancy constant $k$ is well understood. Simulation results have shown that Trickle scales well with network density, suppressing many redundant broadcasts \cite{Levis2008, Levis2004}. In \cite{Meyfroyt2014-4} the authors provide analytical results on Trickle's message overhead and broadcasting rate and show how they depend on $k$ and the network size. These results prove Trickle's scalability and show that its message overhead scales linearly in $k/I_{\max}$. Additionally, ways to approximate the message-count in multi-hop networks are given in \cite{Kermajani2012,  Meyfroyt2014-4}, where the former focuses on random spatial networks, and the latter on grid-like topologies.

However, little is known about the influence of $k$ on other QoS measures such as hop-count, end-to-end delay and load distribution. The authors of \cite{Clausen2013} study the performance of Trickle as a flooding mechanism compared to classic flooding and multipoint relaying. They conclude that while Trickle can outperform both protocols, its performance is highly sensitive to the choice of parameters, stating: \textit{``Simulations showed that the same set of parameters can render Trickle Multicast the best or worst performer in a given scenario"}.

In more recent work \cite{La2014} the authors conclude that flooding using Trickle can perform poorly due to its suppression mechanism. Since $k$ does not change with node density, the suppression mechanism favors nodes with few neighbors, letting them broadcast more often than nodes with more neighbors. This leads to increased traffic along the edges of a network and potentially increased end-to-end delays. They underline the importance of correctly setting $k$ to avoid such issues. Additionally, they write: ``\textit{We had not expected such artifacts - they are rarely mentioned if ever in the literature}". Similar problems have been identified in \cite{Stolikj2015}, where, due to the suppression mechanism, bottleneck topologies have been shown to be prone to extremely large end-to-end delays.

Analytical models for the speed at which Trickle can propagate new data are developed in \cite{Becker2011, Kermajani2014-2, Meyfroyt2014-3}. In \cite{Becker2011} the authors provide a method for deriving the Laplace transform of the distribution function of the end-to-end delay for any network topology.  In \cite{Kermajani2014-2, Meyfroyt2014-3} the authors develop and analyze models for how fast Trickle can propagate data in  chain topologies. The first work considers propagation of data under realistic assumptions, such as the presence of bit-errors, however they limit themselves to sparse networks.  In \cite{Meyfroyt2014-3} the authors assume idealized error-free chain topologies. Additionally, they propose a small extension of Trickle, increasing the speed at which data is propagated. However, all models fail to capture the influence of $k$ on Trickle's performance, due to the complexity of the analysis or by assuming $k=1$.

\subsection{Trickle as a part of RPL}
Next to work focused at analyzing Trickle for flooding, there has also been extensive research on the impact of Trickle parameters on the performance of the RPL routing protocol, which uses Trickle to construct a routing table.

In \cite{Ko2011, Tripathi2010} simulation studies on RPL's performance are presented. However, they are based on earlier versions of the RPL draft, in which Trickle's suppression mechanism is not yet used ($k=\infty$). Both works conclude that in some scenarios the performance of RPL is lacking and additional studies are needed for its usage in large-scale networks.

Later, Trickle's suppression mechanism was deemed necessary to ensure scalability of the protocol and has been made part of RPL's current RFC. Currently, the RFC recommends using $k=10$. Based on this recommendation, additional simulation studies have been performed \cite{Accettura2011, Iova2013}. However, these studies do not consider the effect of topology characteristics on RPL's performance and keep parameter settings fixed.

The authors of \cite{Vallati2013} were the first to consider the effect of the redundancy constant $k$ on RPL's performance. They show that if configured incorrectly, Trickle's suppression mechanism can lead to sub-optimal routes, especially in networks that are heterogeneous in terms of density, such as random spatial topologies. This is again due to the inherent unfairness of Trickle's suppression mechanism. They propose a modification of Trickle, which tries to remove this unfairness by prioritizing nodes that have not broadcasted for a long period of time.

Recently, link instability was identified as a problem for new nodes in a network \cite{Ancillotti2014}. Due to the lack of link quality measurements, new nodes have been observed to blindly connect to the first available node in an RPL network, even though better alternatives might exist. They address this issue by adding a probing phase, where nodes first measure the link quality to their neighbors based on a Trickle timer, before selecting a preferred parent. As a result, nodes take more time to join a network, but benefit from having more stable routes.

Lastly, an extensive simulation study on the effect of the redundancy constant $k$ and $I_{\min}$ on RPL's performance is given in \cite{Kermajani2014}. In their study they consider several network densities and vary $k$ between 1 and 15. They observe that one of the RPL parameters that affects routing table construction to the greatest extent is the redundancy constant $k$. Additionally, they conclude ``\textit{There exists a trade-off between network convergence time and other performance parameters, such as the number of DIO messages transmitted and number of collisions that depends on $k$}" and that setting $k$ should not be done independently of network density.
\section{The redundancy constant}\label{adaptive}
Clearly, the redundancy constant $k$ is one of the most important parameters of Trickle, but its effect on performance is not well understood. The reason that the influence of the redundancy constant is not well studied is twofold. First of all, as was shown in \cite{Kermajani2014, Vallati2013}, the performance of Trickle for a given parameter setting greatly depends on the network topology. This makes studying the effect of the redundancy constant difficult and is why analytical works mostly focus on $k=1$ and simple network topologies. Secondly, as mentioned in the previous section, many of the studies related to Trickle consider it in the context of RPL, where the redundancy constant was only introduced in later draft versions.

Consequently, the current parameter recommendations for configuring Trickle are often vague or conservative. In the Trickle RFC \cite{Trickle-RFC} it is stated that ``\textit{In general, it is much more desirable to set $k$ to a high value (e.g., 5 or 10) than infinity.  Typical values for $k$ are 1-5: these achieve a good balance between redundancy and low cost}". Looking at the RPL RFC \cite{RPL-RFC} we find that the default setting is $k=10$, which is rather arbitrary, stating ``\textit{This configuration is a conservative value for Trickle's suppression mechanism.}" However, in the IETF draft ``Recommendations for Efficient Implementation of RPL" \cite{RPL-recom-draft} we find ``\textit{A constant of 3-5 has been found adequate in deployments.}" Lastly, the latest MPL draft \cite{MPL-draft} recommends using $k=1$, focusing heavily on overhead reduction.

The reason these recommendations are conservative and varying is because optimally setting $k$ is non-trivial and depends greatly on the network topology and the application for which the Trickle algorithm is used. However, one might assume that given some network topology, it should be possible to optimally set the redundancy constant.

Clearly, in sparse networks, where nodes have few neighbors, $k$ should be set to a low value for the suppression mechanism to work. In very dense networks, a higher value of $k$ makes more sense, since nodes have to compete with many other nodes for the medium, and a low value of $k$ can thus lead to sub-optimal data dissemination. Hence, when setting $k$, it is important to take network density into account. However, in many cases, networks are heterogeneous and consist of both sparse and dense parts, which makes setting $k$ difficult.

\subsection*{\textit{Adaptive-}$k$: a density-aware redundancy constant}
With these considerations in mind, it makes sense to set $k$ for each node individually. Ideally, we would want nodes to set their own $k$ in a distributed fashion according to their node degree. However, since we want Trickle to be self-sufficient, we cannot rely on nodes having perfect knowledge on the number of neighbors they have. What we do know is that nodes keep track of the number of Trickle messages they receive during an interval with a counter $c$. This $c$ contains implicit information on the number of neighbors of that node. Therefore, we propose an extension to the Trickle algorithm, called \textit{adaptive-}$k$, which leverages this information and allows nodes to set their value of $k$ autonomously. This extension is done by a slight modification of rule 4 of the algorithm:
\begin{framed}
\begin{itemize}
\item[$4^*.$] When a node's timer hits time $I$, it sets $k$ equal to $f(c)$, it increases its interval length to $\min(2I, I_{\max})$ and starts a new interval.
\end{itemize}
\end{framed}
Here $f$ is some predefined function, which is the same for all the nodes of the network. We argue that a natural candidate is the following function:
\begin{equation}\label{f}
f(c)=\begin{cases}
k_{\min}, &\alpha c < k_{\min},\\
\lfloor\alpha c\rfloor, &k_{\min} \leq \alpha c \leq k_{\max},\\
k_{\max}, & \alpha c > k_{\max},
\end{cases}
\end{equation}
with some fixed $\alpha\in[0,1]$ and $k_{\min}, k_{\max}\in\mathbb{N}$. The function $f$ should be bounded by below by some $k_{\min}$ to avoid a deadlock with all nodes having $k=0$. One should think of $k_{\min}$ being small, i.e. 1 or 2. Throughout this paper we assume $k_{\min}=1$. Furthermore, $f$ should be bounded from above by $k_{\max}$ to ensure scalability of the algorithm. Lastly, as recommended in the Trickle RFC, $k_{\max}\times I_{\min}$ should at least be two to three times as long as it takes to transmit $k_{\max}$ packets.

Intuitively, this extension does what we would like it to do. Whenever a node receives many broadcasts during an interval, it knows it has many neighbors, and hence it should have a high $k$ value in order to be able to compete for the medium. When a node receives few transmissions, it either does not have a lot of neighbors, or its neighbors are having a hard time broadcasting their own information, and for both cases the node should lower its redundancy constant $k$.

Note that our extension is backward compatible with the Trickle RFC: the RFC itself acknowledges that nodes can be configured with a different redundancy constant, with the possible drawback of an uneven load distribution. In the next section we show that Trickle with  \textit{adaptive-}$k$ actually leads to a more even load distribution.

\section{Evaluation of Trickle with adaptive-k}
\label{analysis}
In this section we will analyze the Trickle algorithm with \textit{adaptive-}$k$ for some special network topologies, highlighting its benefits. For simplicity, we assume in all these cases that all nodes are perfect receivers and transmitters, meaning that transmissions never fail and are received instantaneously, so that we can focus on the performance of Trickle without considering any MAC-layer protocols. Furthermore, we focus only on the control traffic generated by Trickle and assume all nodes to have the same data and that $I=I_{\max}$ for all nodes.

\subsection{Single-cell network}
First, let us consider a simple network consisting of $n$ nodes that are all within communication distance of each other. Let $f(c)$ be as in Equation \eqref{f} with $k_{\min}=1$.

Assume all nodes are synchronized, i.e. all nodes start new intervals at the same time. Then it is easy to see that, if $\alpha<1$, regardless of their initial value of $k$, each interval nodes will decrease their $k$, until eventually they all set $k=1$. From that point on nodes will never receive more than one transmission per interval, and hence $k$ will stay fixed to one. For most applications this setting can be regarded as optimal, since we only need one transmission to reach all nodes.

Now assume the intervals of nodes are not necessarily synchronized. If $\hat{k}$ is the maximum value for the redundancy constant among nodes during a single interval, then we know that each interval there will be less than $2\hat{k}$ transmissions \cite{Meyfroyt2014-4}. Hence, if $\alpha\leq 1/2$, each interval nodes will decrease their $k$ and eventually all set $k=1$. From that point on, there will be at most $2$ transmissions per interval. If, however, $\alpha>1/2$ and $n\rightarrow \infty$, nodes will eventually set $k=k_{\max}$ and there will be at most $2k_{\max}$ transmissions per interval. This also highlights the fact that $f$ should be bounded to ensure scalability. However, note that regardless of $\alpha$ and $k_{\min}$, Trickle with \textit{adaptive-}$k$ will have at most as many transmissions per interval as the original Trickle with $k=k_{\max}$.

\begin{figure}[!h]
\begin{center}
  \includegraphics[width=.6\linewidth]{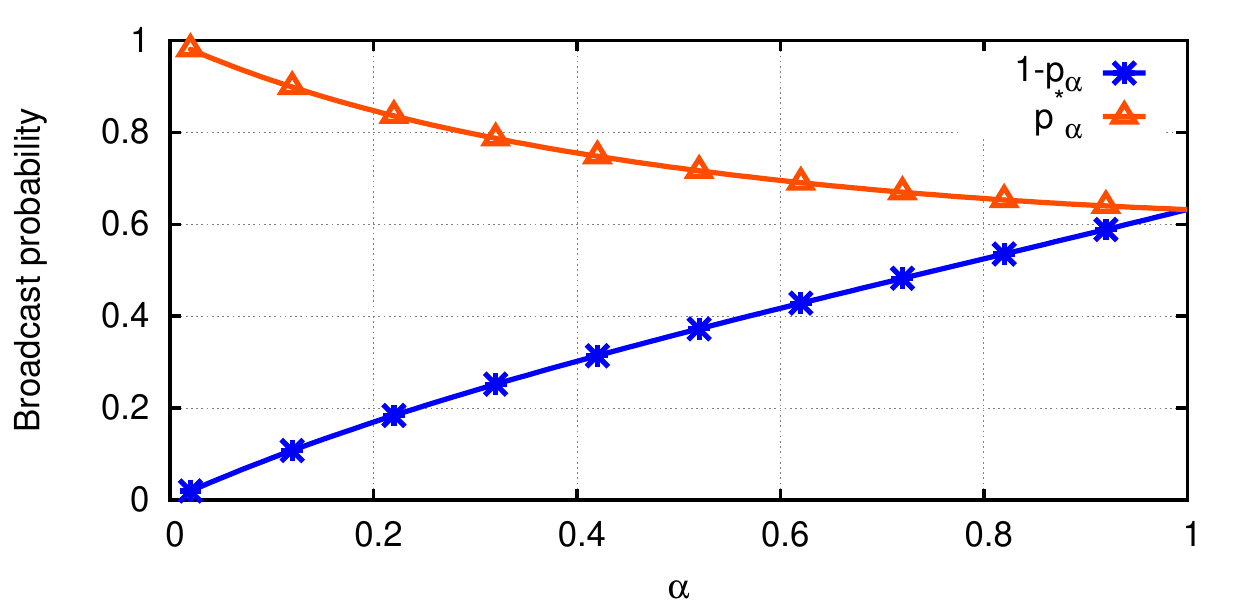}
  \caption{Asymptotic broadcasting probabilities of \emph{adaptive-}$k$ in the star network. $1-p_{\alpha}$ is the probability that the central node broadcasts, while $p^*_{\alpha}$ is the probability that other nodes broadcasts.}
  \label{figp}
  \vspace{-1.0em}
  \end{center}
\end{figure}

\begin{figure*}
\begin{center}
  \subfloat[Sparse network (average degree $ =5$)]{
  \label{fig:sparse1}
  \includegraphics[width=0.33\linewidth]{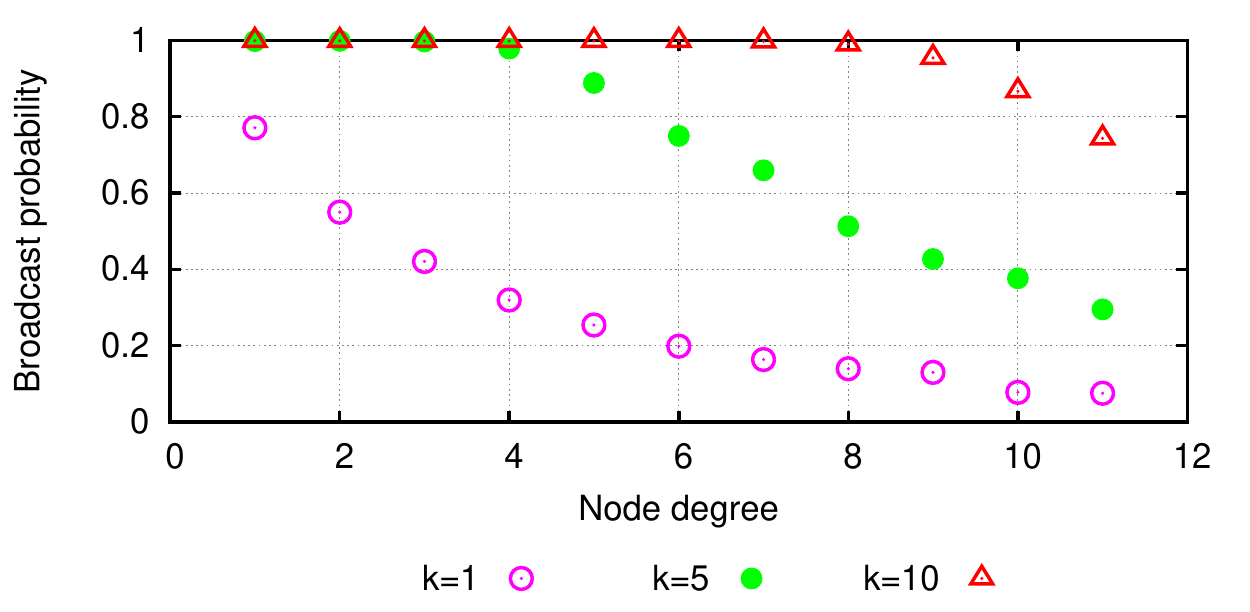}
}
 \subfloat[Medium network (average degree $ =10$)]{
  \label{fig:medium1}
  \includegraphics[width=0.33\linewidth]{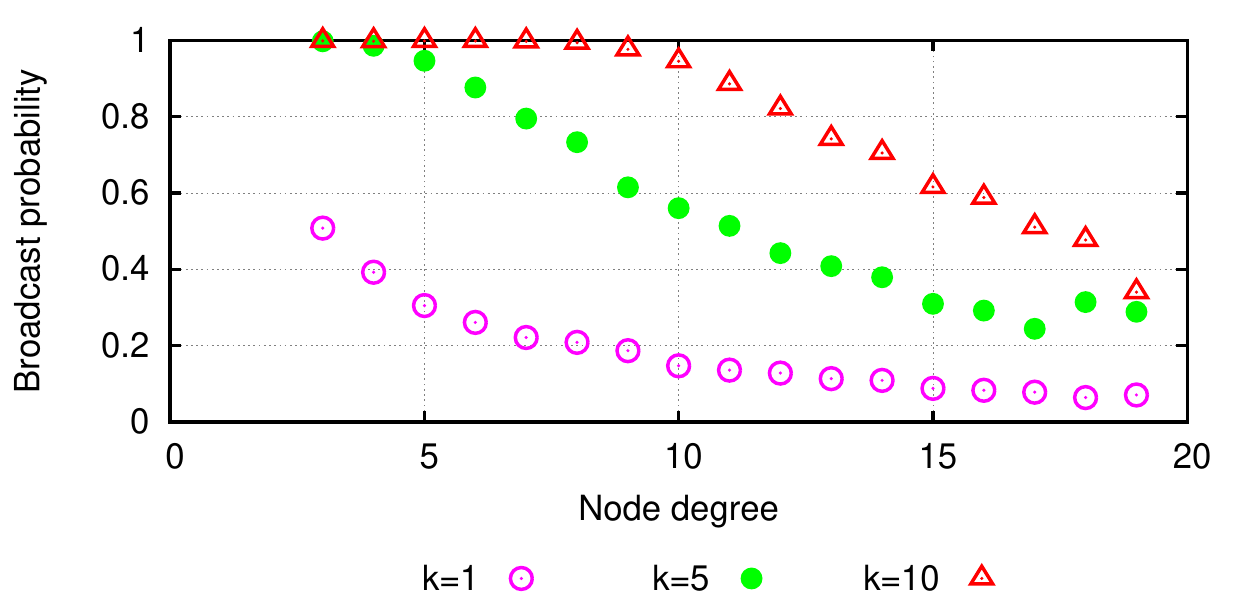}
}
 \subfloat[Dense network (average degree $ =15$)]{
  \label{fig:dense1}
  \includegraphics[width=0.33\linewidth]{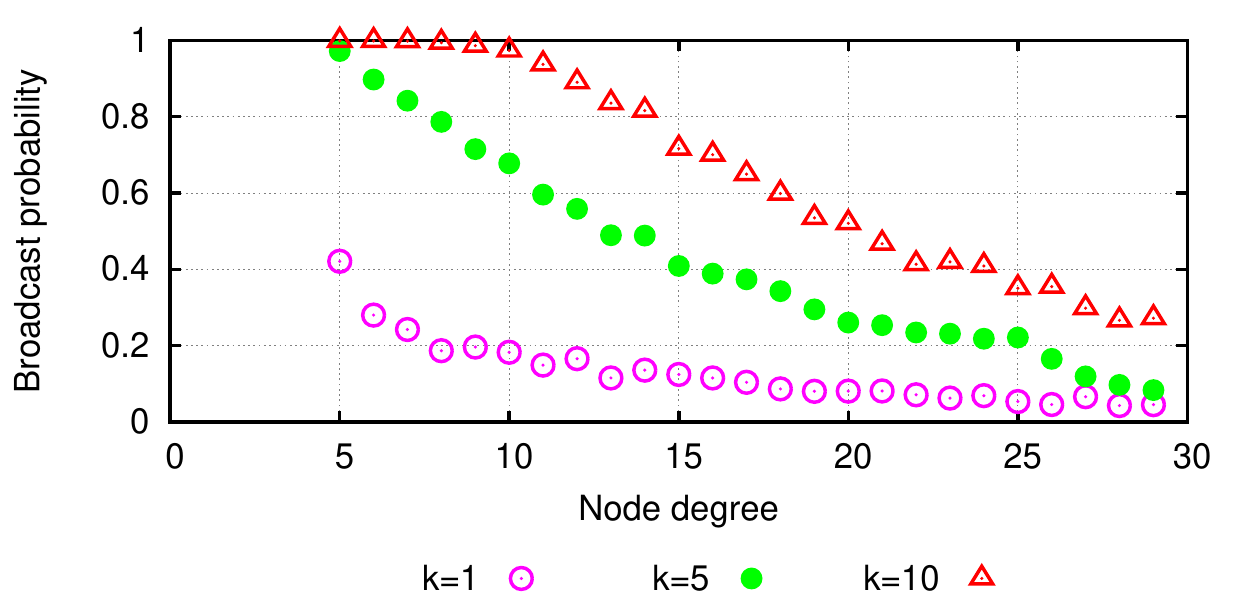}
}
\end{center}
  \caption{Broadcasting probability per node degree for the \textit{original} Trickle algorithm. The three figures correspond to the three different network densities.}  \label{fig1}
  \vspace{-1.0em}
\end{figure*}
\begin{figure*}
\begin{center}
  \subfloat[Sparse network (average degree $ =5$)]{
  \label{fig:sparse2}
  \includegraphics[width=0.33\linewidth]{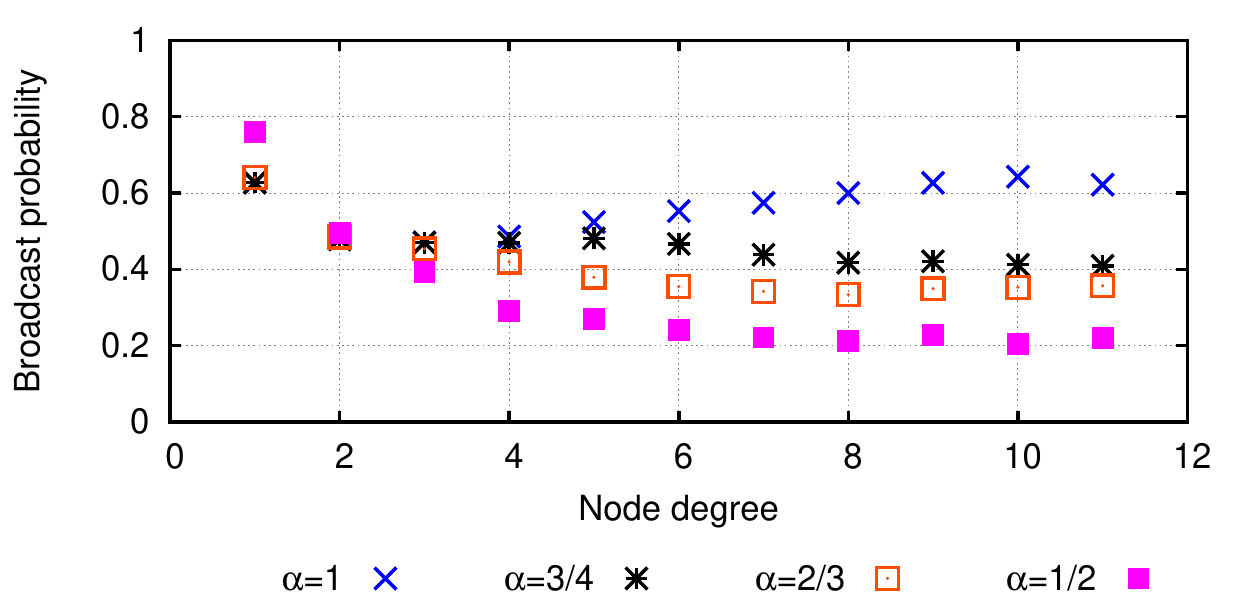}
}
 \subfloat[Medium network (average degree $ =10$)]{
  \label{fig:medium2}
  \includegraphics[width=0.33\linewidth]{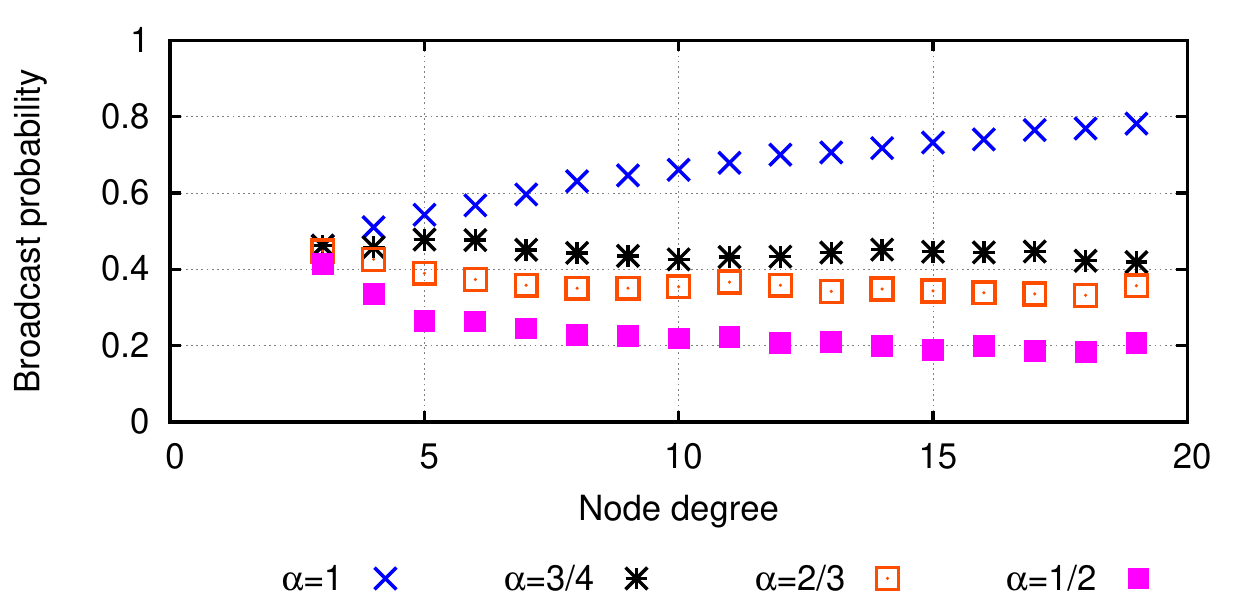}
}
 \subfloat[Dense network (average degree $ =15$)]{
  \label{fig:dense2}
  \includegraphics[width=0.33\linewidth]{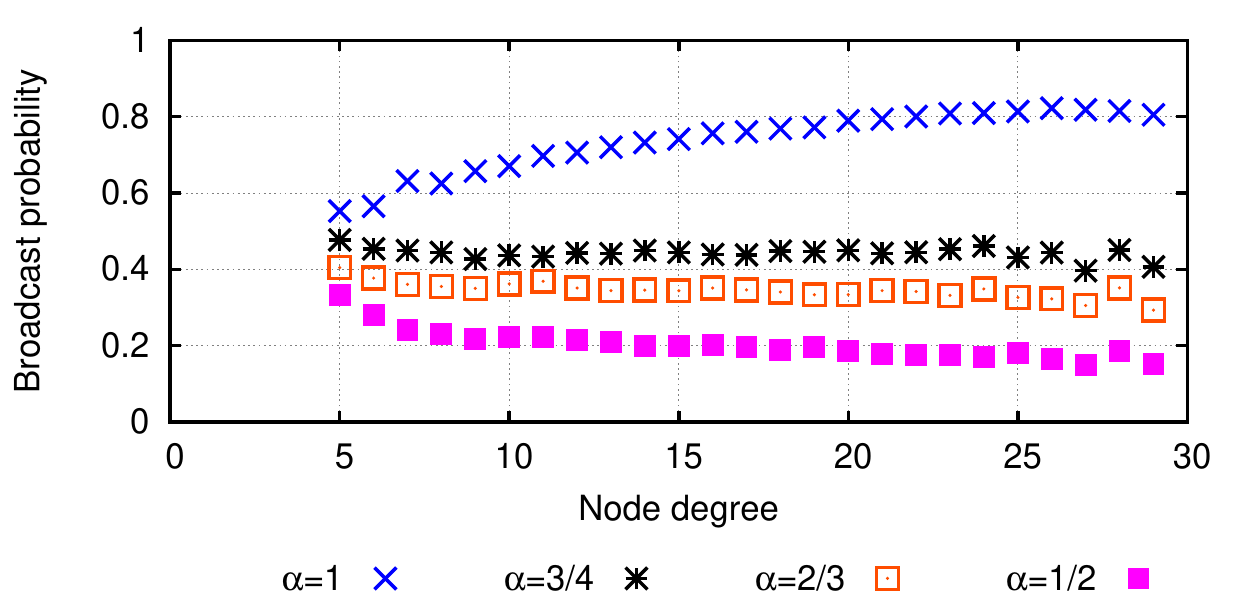}
}
\end{center}
  \caption{Broadcasting probability per node degree for the \textit{adaptive-}$k$ Trickle algorithm. The three figures correspond to the three different network densities.}  \label{fig2}
\end{figure*}

\subsection{Star network}
Consider now a network consisting of $n+1$ nodes, where one central node is connected to all other $n$ nodes, which are all not connected to any other nodes. Note that this is in contrast with the previous scenario, where all nodes had the same number of neighbors. Assume all nodes are synchronized, i.e. all nodes start intervals at the same time.

Suppose the unmodified Trickle algorithm is used with $k<n+1$, then the central node will broadcast if it is among the first $k$ nodes to schedule a broadcast, which happens with probability $\frac{k}{n+1}$. Hence, if $k=1$, the central node will broadcast a fraction $\frac{1}{n+1}$ of the intervals and all other nodes will broadcast a fraction $\frac{n}{n+1}$ of the intervals. This clearly results in a very unfair load distribution and $\frac{n^2+1}{n+1}$ broadcasts per interval on average. If $k>1$, the central node will broadcast a fraction $\frac{k}{n+1}$ of the intervals and the rest will always broadcast, also leading to an unfair load distribution and a high message count.

Now suppose the modified Trickle algorithm is used with $f(c)$ as in Equation \eqref{f}, but this time, for illustrative purposes, let $k_{\max}=\infty$. It is easy to see that eventually, only the central node will be adapting its redundancy constant and the rest of the nodes will set $k=1$. Suppose the central node starts an interval with $k=\alpha n$ (i.e. it heard all of its neighbors transmit the previous interval), and suppose it is the $m$'th node to schedule a broadcast during that interval. If $m\geq \alpha n$, then the central node's message will be suppressed and $k=\alpha n$ the next interval as well. If $m<\alpha n$, then the central node will broadcast, suppressing the transmissions of the last $n-m$ nodes to schedule a broadcast, and hence $k=\alpha(m-1)$ the next interval. Continuing this reasoning one can deduce that the redundancy constant of the central node evolves according to a Markov chain. Analyzing the Markov chain for $n\rightarrow \infty$ (see Appendix) then tells us the following.

Let $p_{\alpha}$ be the probability that the central node's broadcast is suppressed during an arbitrary interval. Then, as $n\rightarrow \infty$,
\begin{equation} p_\alpha=\left(\sum_{i=0}^\infty \frac{\alpha^{i(i+1)/2}}{i!}\right)^{-1}.\end{equation}
Note that the central node successfully broadcasts with probability $1-p_{\alpha}$.
Let $p^*_{\alpha}$ be the probability that a node other than the central node successfully broadcasts during an arbitrary interval. Then, as $n\rightarrow \infty$,
\begin{equation} p^*_\alpha=\frac{1}{\alpha}(1-p_{\alpha}).\end{equation}
Plots of the probabilities are shown in Figure \ref{figp}.

\noindent
We conclude that for $\alpha=1$ the star network with dynamic $k$ is asymptotically fair and
\begin{equation}p^*_1=1-p_1=1-\frac{1}{e}.\end{equation}

Note that this is quite remarkable, since as $n\rightarrow \infty$, the central node has to compete against infinitely many other nodes for the medium. Moreover, on average Trickle with \textit{adaptive-}$k$ will have fewer broadcasts per interval than the original algorithm.

\subsection{Random spatial network}
We now consider random spatial networks consisting of $200$ nodes placed uniformly in a square region. Since analyzing this case analytically is too difficult, we resort to simulations. We consider the original Trickle algorithm with $k=1, 5$ and $10$, as well as the Trickle algorithm with \textit{adaptive-}$k$ for four different functions $f(c)$ of the form as in Equation \eqref{f}.

From the single-cell network we learned that for dense homogeneous networks setting $\alpha$ to a lower value is smart, since it reduces the amount of traffic. On the other hand, from the star network we learned that for highly heterogeneous networks, a high value of $\alpha$ contributes to more fairness in the network, which can also lead to a reduction of the amount of traffic. Since realistic spatial networks should be a mix of the two scenarios, we will consider several values of $\alpha$ in the range $[\frac{1}{2},1]$, and try to determine what value performs well for such topologies. For this purpose we set $k_{\min}=1$ and $k_{\max}=30$, to be able to capture the effect of the number of neighbors of a node on its message count. Additionally, we consider three network densities, where the average node degree is either 5, 10 or 15. For each setting and density, we simulate 10 networks with different topologies for 200 time units, where all nodes start with $I=I_{\max}=1$ and uptodate information. For each node degree we calculate the average broadcasting probability, i.e., the fraction of intervals nodes with that number of neighbors broadcast.

Results for the original Trickle algorithm can be found in Figure \ref{fig1}. We can immediately see Trickle's tendency to favor low-degree nodes, letting them broadcast often. Nodes with degree bigger than $k$ have a hard time transmitting. This effect is clear for each of the three network densities.

\begin{figure}[!h]
\begin{center}
  \includegraphics[width=.6\linewidth]{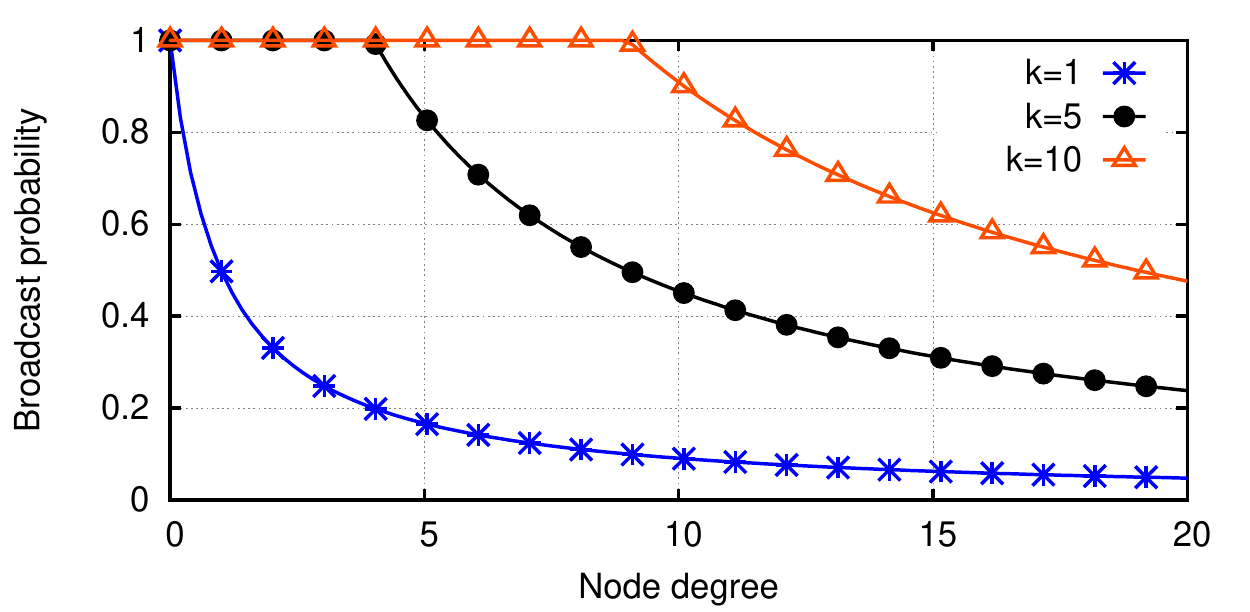}
  \caption{Estimate for the broadcasting probabilities for different values of $k$.}
  \label{figbpk}
  \end{center}
\end{figure}

The trend in these probabilities is also as one would expect. If a node has less than $k$ neighbors, it will broadcast almost every interval. If a node however has $N\geq k$ neighbors then, ignoring the fact that neighbors can be suppressed as well, it will broadcast if it is among the first $k$ nodes to schedule a broadcast during that interval, which roughly happens with probability $k/(N+1)$. Hence, the probability of a node with $N$ neighbors broadcasting can roughly be estimated by $\min(1,k/(N+1))$. In Figure \ref{figbpk} we have plotted this probability for $k=1$, $5$ and $10$ and one can  see the same trends in the plots of Figure \ref{fig1}.

If we then look at the results for the Trickle algorithm with \textit{adaptive-}$k$ in Figure \ref{fig2}, we find that transmission load is being distributed more fairly. Note, however, that for $\alpha=1$, high degree nodes are being favored and all nodes broadcast with relatively high probability. For $\alpha=1/2$, we see there is still a tendency to favor low degree nodes, especially in sparse networks, but less so than for the original Trickle algorithm. The most appropriate setting for $\alpha$ for these networks in terms of fairness seems to be in the range of $2/3$ and $3/4$. For these two settings on average there are less broadcasts than for $k=5$, while the transmission load is distributed fairly.

\begin{figure}[h!]
\center
  \includegraphics[width=0.7\linewidth]{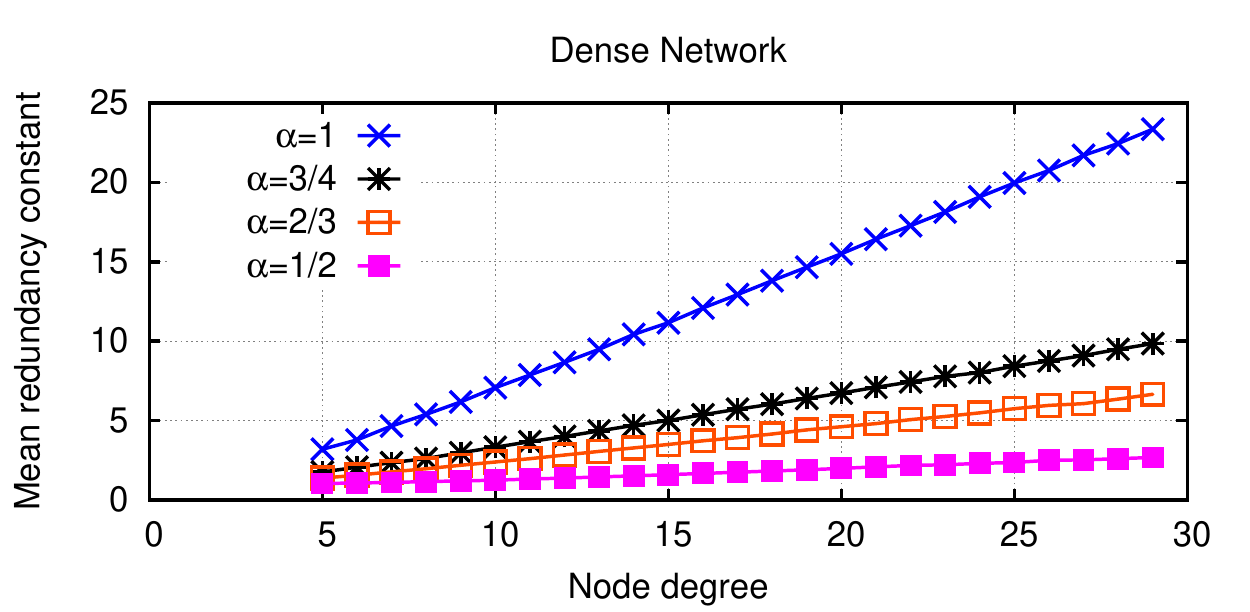}
  \caption{The average redundancy constant per node degree.}\label{fig3}
\end{figure}

Finally, in Figure \ref{fig3}, we show the average value of $k$ per node degree for the dense network scenario. Clearly, there is a linear dependency on the number of neighbors a node has and its average redundancy constant, where the slope depends on $\alpha$. This is also what one would like to see; the more neighbors a node has, the higher its redundancy constant should be to be able to compete for the medium.

\section{A case study: RPL}\label{RPL}
Lastly, to confirm that the \textit{adaptive-}$k$ extension to the Trickle algorithm can improve the performance of the protocols that it is used in, we perform a case study using RPL. First, we provide a general description of the RPL routing protocol. Then, we present simulation results for network of varying density using RPL. Finally, we confirm the simulation results with experimental data from a physical test bed.

\subsection{RPL basics}
RPL is a distance vector routing protocol for low-power and lossy networks (LLNs) that uses the Trickle algorithm to build a Destination Oriented Directed Acyclic Graph (DODAG). The DODAG is a tree-like network-graph, rooted in a single node, in which all nodes learn a route towards the root node. While RPL supports both upwards and downwards routes, in this work we only focus on upward routes. The DODAG is shaped according to one or more objective functions (OF), which can specify metrics for the cost of routes or give rules/constraints when building links.

The DODAG is built starting from the root as follows. The root advertises information about the graph using DODAG Information Object messages (DIO), which are disseminated based on a Trickle timer. DIO messages contain information about the DODAG, its configuration parameters and the `rank' of the sender in the DODAG - a monotonically decreasing measurement indicating the distance of the sender to the root according to the objective function(s). Non-root nodes process these DIOs and based on the objective function and/or some local criteria, decide whether to join the network. After joining, they establish which directly reachable nodes can forward data most efficiently towards the root (i.e. have the lowest rank), and select one of them as the \emph{preferred parent}. Then, they compute their own rank, and start transmitting DIO messages.

Once all nodes have selected a parent and have become a part of the DODAG, we say that the DODAG has been formed. Whenever a node then needs to send a message to the root, it sends it to its parent, which then forwards it to its own parent until it reaches the root.

Clearly, in order to be able to forward messages as efficiently as possible, nodes need to quickly update their rank to the `optimal' rank. However, as we saw in the previous section, the original Trickle algorithm tends to favor nodes with low degree. Therefore, low degree nodes will be able to broadcast DIO's more often than high degree nodes. This introduces the problem that nodes tend to favor low degree nodes as their parents, possibly leading to sub-optimal routes, since routes through high degree nodes might be more efficient. One hopes that using Trickle with \textit{adaptive-}$k$ could solve this problem, since it distributes the transmission load more evenly.

\begin{figure}[!ht]
\begin{center}
  \includegraphics[width=0.5\linewidth]{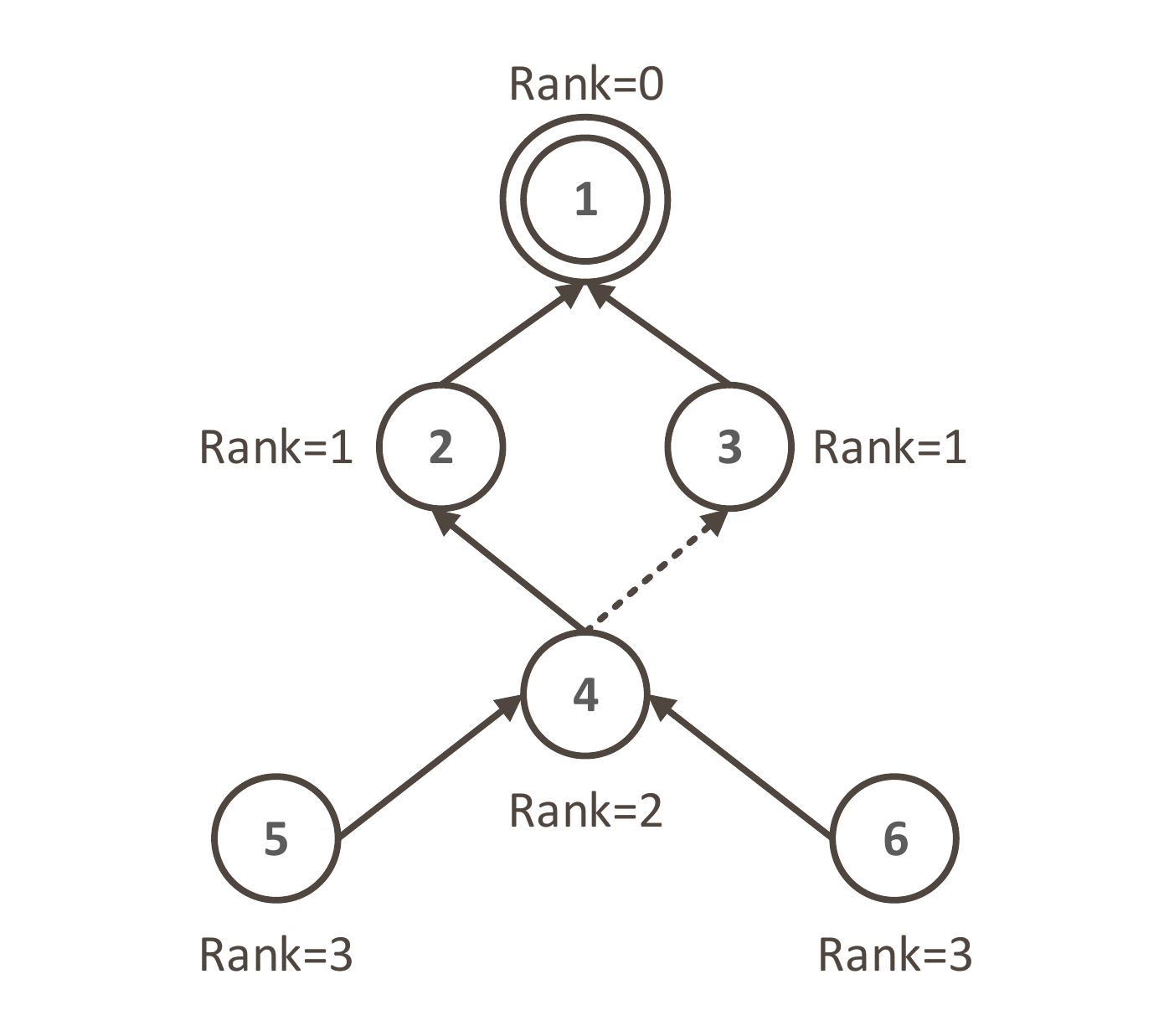}
  \caption{Example of a simple DODAG using hop count as an Objective Function. Node 4 has two parents, and selects node 2 as the preferred one.}
  \label{dodag}
  \vspace{-1.0em}
\end{center}
\end{figure}
\begin{figure*}
\begin{center}
  \subfloat[DODAG formation time]{
  \includegraphics[width=.33\linewidth]{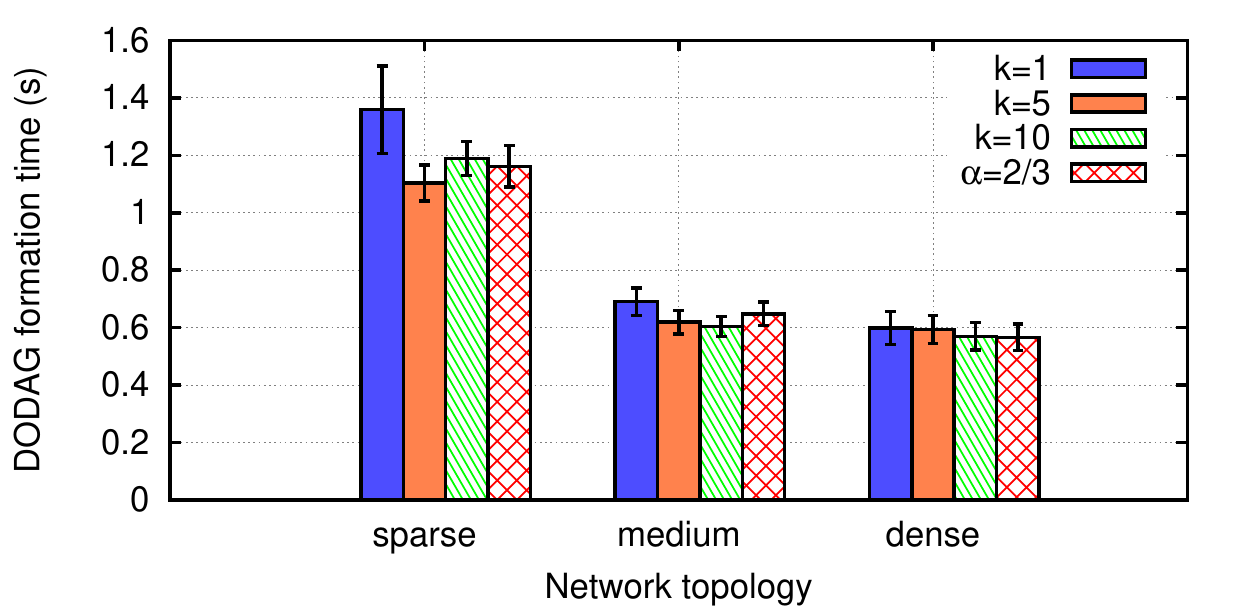}
  \label{formation}
}
  \subfloat[Average number of DIO's]{
  \includegraphics[width=.33\linewidth]{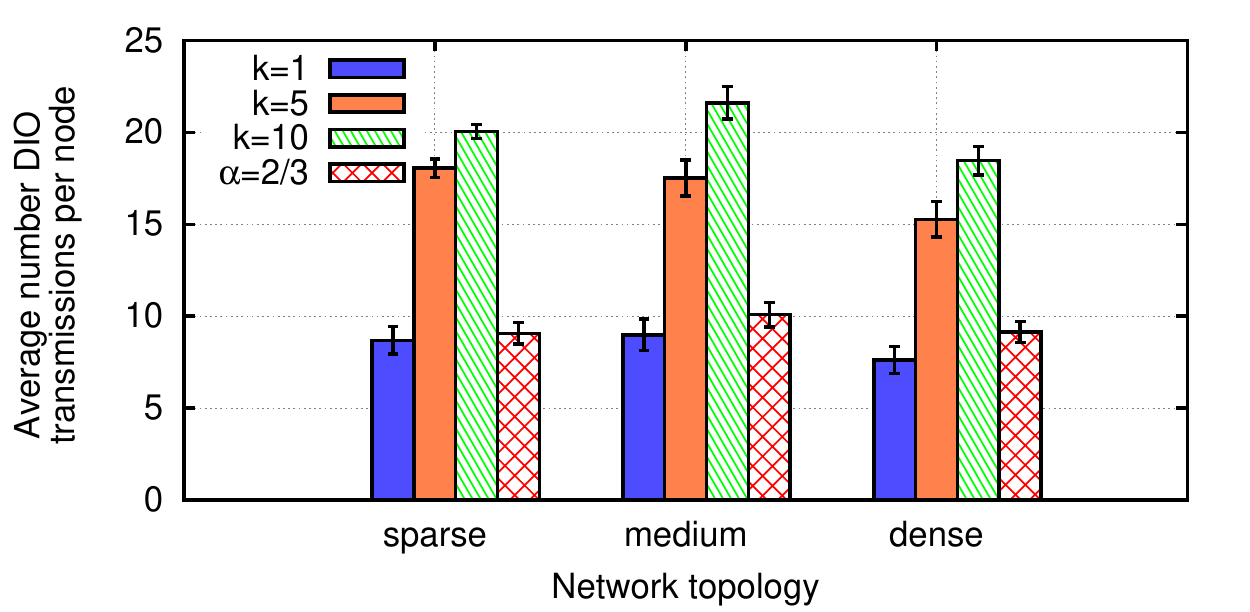}
  \label{dios}
}
  \subfloat[Network stretch]{
  \includegraphics[width=.33\linewidth]{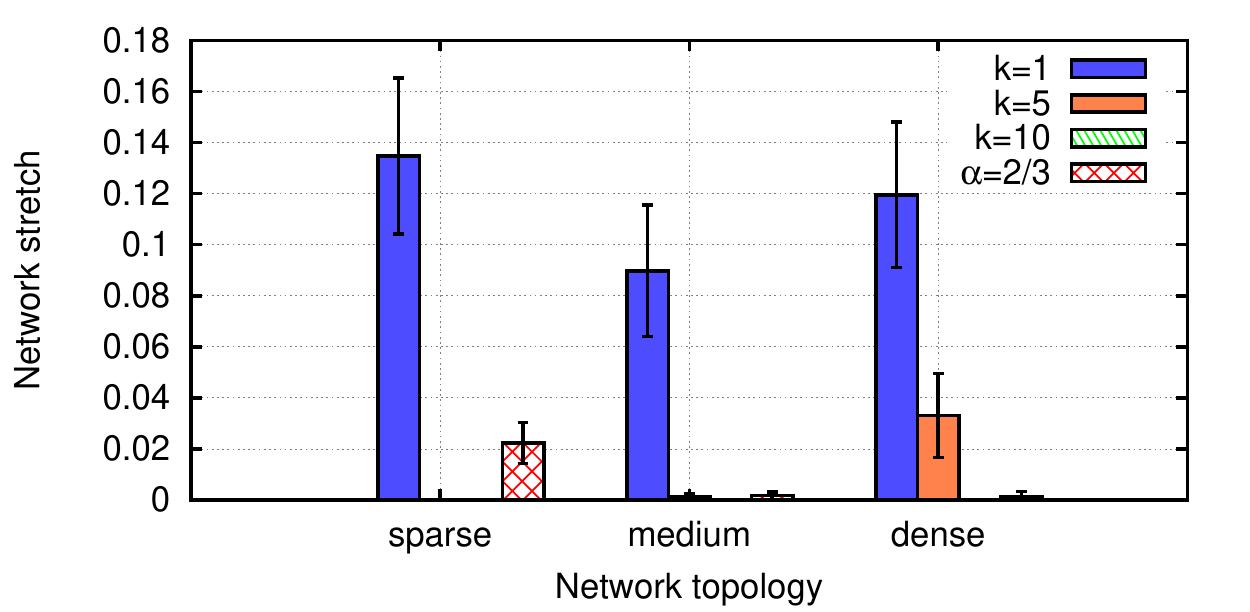}
  \label{stretch}
}
\end{center}
\caption{Influence of the redundancy constant in RPL for different topologies. All values show the average of 100 runs and the 95\% confidence interval.}
\label{simulations}
\end{figure*}
\subsection{Simulation results}
We implement \textit{adaptive-}$k$ as part of the Contiki 2.7 operating system~\cite{Dunkels04}. Contiki is an open-source operating system for constrained devices, which includes a full 6LoWPAN stack, together with an implementation of the RPL protocol, called ContikiRPL. We simulate different topologies in Cooja, a cross-level simulator for Contiki. Cooja~\cite{Osterlind2006} internally uses the MSPsim device emulator for cycle accurate Tmote Sky emulation, as well as a symbol accurate emulation of the CC2420 radio chip. We use the Unit Disk Graph Radio Medium (UDGM) model for radio propagation, with no loss. UDGM penalizes collisions heavily, while end-to-end delivery fails only due to filled MAC queues. At the link layer, we use the default CSMA/CA implementation in Contiki with no radio duty cycling (nullrdc). The RPL DODAG is formed according to the Objective Function Zero (OF0)~\cite{OF0-RFC}. OF0 in Contiki is implemented as a hop-count based selection metric, which uses local expected transmission count (ETX) measurements to select between parents with the same rank. Unnecessary parent switches are avoided by adding a simple hysteresis mechanism~\cite{MRHOF-RFC}.

\begin{figure}
\begin{center}
	\includegraphics[width=0.33\textwidth]{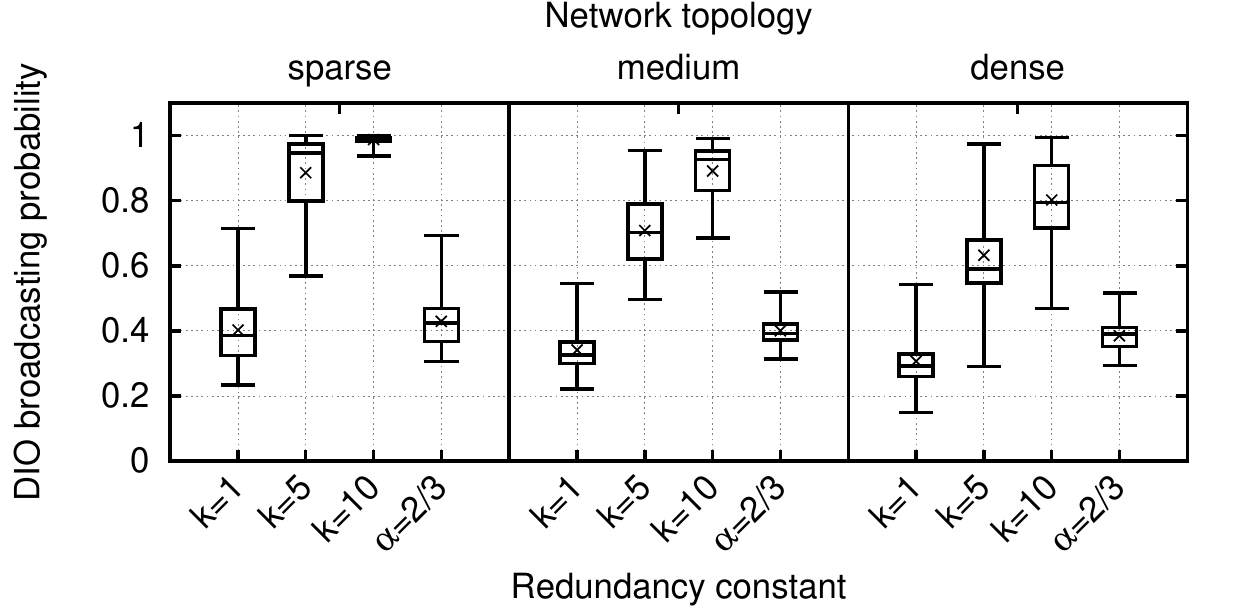}
  \caption{Cumulative statistics on the average DIO broadcasting probability per node for different network densities. The whiskers show the minimum/maximum value, the start and end of the box show the 25th/75th percentile, the line in the box is the median, and the cross is the mean.}
  \label{rplprob}
  \vspace{-1.5em}
\end{center}
\end{figure}

As in the previous section, we again look at topologies with three different network densities, where this time one root and 100 non-root nodes are uniformly placed in a square area of 100$\times$100 meters. For each topology, the transmission range is such that the average node degree is 5, 10 and 15, respectively. We simulate a Constant Bit Rate data gathering application - every non-root node sends one 80-byte packet (including all headers) to the root node every minute. For each topology, we consider the original Trickle algorithm with $k=1,5$ and $10$ and \textit{adaptive-}$k$ Trickle with $\alpha=\frac{2}{3}$, $k_{\min}=1$ and $k_{\max}=10$. We simulate each configuration 100 times for 2 hours with $I_{min}=2^3ms$ and $I_{\max}=2^{23}ms$. We measure the mean time until formation of the first DODAG, the mean number of DIO transmissions per node and the mean network stretch after 2 hours. Network stretch is defined as the fraction of nodes with a rank higher than the minimal rank, i.e. take more hops than necessary to reach the root.

%\begin{figure}[!h]
%\begin{center}
%  \includegraphics[width=\linewidth]{time-dodag}
%  \caption{DODAG formation time}\label{formation}
%    \end{center}
%\end{figure}
%\begin{figure}[!h]
%\begin{center}
%  \includegraphics[width=\linewidth]{dio-total}
%  \caption{Average number of DIO's}\label{dios}
%  \end{center}
%\end{figure}
%\begin{figure}[!h]
%\begin{center}
%  \includegraphics[width=\linewidth]{ns-end}
%  \caption{Network stretch}\label{stretch}
%  \end{center}
%\end{figure}

First of all, we find that DODAG formation time is not greatly affected by the choice of $k$; only for sparse networks does the formation time suffer from low values of $k$ (Figure~\ref{formation}). Secondly, for the original Trickle algorithm, the average number of DIO's increases quickly with $k$, as expected. However, the DIO count of Trickle with \textit{adaptive-}$k$ is comparable to that of the original algorithm with $k=1$ (Figure~\ref{dios}). Lastly, for the original Trickle algorithm with low $k$ the average network stretch remains high, probably due to Trickle favoring low degree nodes. As $k$ increases the network stretch decreases; nodes are able to broadcast more easily, allowing for discovery of better routes at the cost of high overhead. However, we find that for \textit{adaptive-}$k$, for every scenario the network stretch after 2 hours is almost zero; only in the sparse case there are 2 or 3 nodes that have not yet discovered the optimal route (Figure~\ref{stretch}), which could potentially be avoided by increasing $k_{\min}$.

Furthermore, the DIO broadcasting probability per node is similar as in the random spatial simulations, with \textit{adaptive-}$k$ distributing the load most fairly, while still suppressing many transmissions (Figure~\ref{rplprob}). 

In summary, the network stretch shows that Trickle with \textit{adaptive-}$k$ allows for good routes to be discovered, as if $k$ was high, while only broadcasting few DIO's, as if $k$ was set low, while distributing the message load evenly among nodes.  

\subsection{Experimental results}
To confirm the simulation results, we implement the same code as in the previous section, on a set of 43 WSN430 nodes in the Rennes IoT-Lab physical test bed. WSN430 nodes have the same MSP430 micro-controller and TI CC2420 radio chip as the Tmote Sky. We configure the nodes to use the minimum available transmission power (-25 dBm), which is enough to form a network with at most 4-hops to the root. To enable more accurate measurement, we use larger Trickle intervals, $I_{min}=2^4ms$ and $I_{\max}=2^{24}ms$. The rest of the parameter settings are identical as in the simulations. For each value of the redundancy constant, we perform five experiments, where each experiment runs for one hour. As nodes are booting up randomly, links are fairly unstable, and there is no central clock in the system, we only show the results for the overall number of transmissions as measured at the link-layer, and the end-to-end packet delivery ratio. All charts show the average measured values and the 95\% confidence interval of the mean.
\begin{figure}[h!]
\begin{center}
\vspace{-0.7em}
	\includegraphics[width=0.33\textwidth]{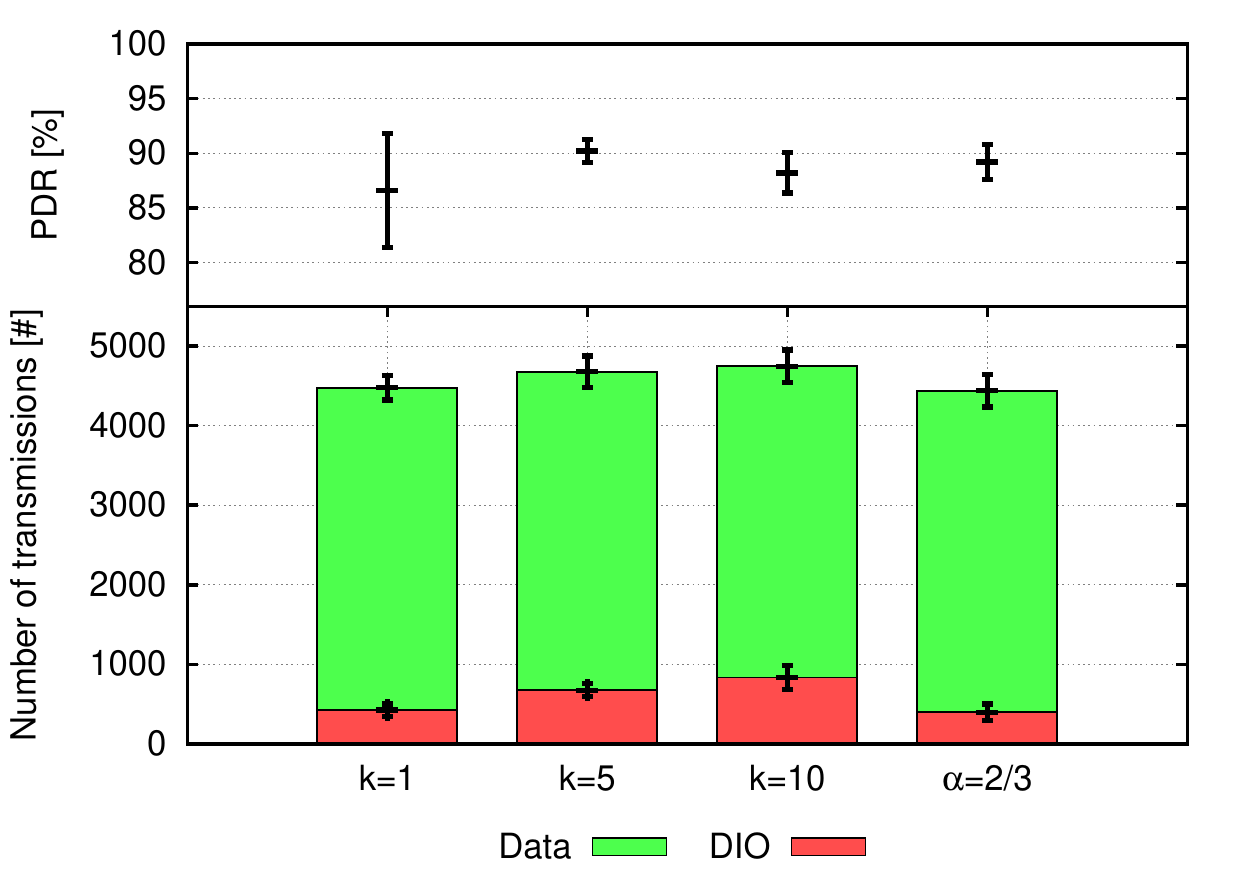}
	\vspace{-0.5em}
  \caption{Total number of DIO transmissions, data transmissions and end-to-end packet delivery ratio (PDR) during 1h of operation at the IoT-Lab test bed.}
  \label{iotlabtx}
  	\vspace{-0.7em}
\end{center}
\end{figure}

The experimental results confirm the simulation results; setting a proper value of the redundancy constant significantly impacts the overall traffic in the network (Figure~\ref{iotlabtx}). In this particular network topology, a high value for the redundancy constant does not improve routing, and therefore should be set low. The proposed \textit{adaptive-}$k$ handles the situation well, balancing between the overhead control traffic and high end-to-end packet delivery ratio.

\section{Conclusions}\label{conclusion}
In this paper we studied the effect of the suppression mechanism on the performance of the Trickle algorithm. The current Trickle standard proposes a fixed value of the redundancy constant, which is sub-optimal in networks of varying density. As a result, there is no clear agreement within the research community to what the preferred setting should be.

By looking at different network topologies, we showed that having a fixed value for the redundancy constant makes Trickle unfair in terms of load distribution, favoring nodes with few neighbors. Moreover, it makes Trickle vulnerable to changing network topologies. Depending on the application where Trickle is used, this unfairness can lead to various consequences, such as increased end-to-end delays (MPL) or creation of sub-optimal routes (RPL).

As a solution, we proposed \textit{adaptive-}$k$, an extension of Trickle which allows nodes to set their own redundancy constant according to local information on network density. Through analysis and simulations we showed that this extension makes Trickle more fair in terms of load distribution, while still suppressing many redundant transmissions. Finally, by simulations and experiments on a physical test bed, we showed that the \textit{adaptive-}$k$ extension improves the performance of the RPL routing protocol, by keeping the overhead of control traffic low, while still discovering good routes.

As future work, we plan to investigate the performance of other possible functions $f(c)$ for \textit{adaptive-}$k$. Additionally, we aim to extend the analysis to the minimum Trickle interval. The Trickle RFC already gives guidelines that is should be set with respect to the redundancy constant, but it greatly depends on the physical deployment of the networks. We believe that having both settings dynamic, depending on the network topology, would give better out-of-the box performance.

\section*{Acknowledgments}
The authors would like to thank Onno Boxma for valuable discussions on the mathematical models and analysis presented in this work. This work is suported in part by the Dutch P08 SenSafety project, as part of the COMMIT program.



%\bibliographystyle{IEEEtran}

\bibliographystyle{IEEEtran}
\bibliography{sigproc}  % sigproc.bib is the name of the Bibliography in this case
% ACM needs 'a single self-contained file'!
% APPENDICES are optional

\appendix[Asymptotic analysis star network]
%Appendix A
Denote the Markov chain of the redundancy constant of the central node by $\boldsymbol{K^{(n,\alpha)}}=\{K^{(n,\alpha)}_i\}_{i=0}^\infty$, i.e. the central node's $k=K^{(n,\alpha)}_i$ in the $i$th interval. First consider the case $\alpha=1$. Let $\mathcal{S}=\{0, 1, 2,\text{ ..., }n\}$ denote the state space of the chain,
where the first two states correspond to the central node having $k=1$ and receiving zero or one transmissions during the previous interval respectively, and the other states correspond to the current value for $k$. Then the Markov chain with state space $\mathcal{S}$ has the following transition matrix
\[\small P=\begin{pmatrix}
\frac{1}{n+1} & 0  & \cdots & 0 & 0 & 1-\frac{1}{n+1}\\[5pt]
\frac{1}{n+1} & 0  & \cdots & 0 & 0 & 1-\frac{1}{n+1}\\[5pt]
\frac{1}{n+1} & \frac{1}{n+1} & \cdots  & 0 & 0 & 1-\frac{2}{n+1}\\[5pt]
\vdots & & \ddots & & & \vdots\\[5pt]
\frac{1}{n+1} & \frac{1}{n+1} & \cdots & \frac{1}{n+1}  &0 & 1-\frac{n-1}{n+1}\\[5pt]
\frac{1}{n+1} & \frac{1}{n+1} & \cdots & \frac{1}{n+1} & \frac{1}{n+1} & 1-\frac{n}{n+1}
\end{pmatrix}.
\]
With some calculus one can use $P$ to determine the steady-state distribution $\{q_j\}_{j=0}^n$ of the Markov chain. This allows us to determine the steady-state probability that a broadcast by the central node is suppressed when $n$ grows large, that is
\begin{equation}\label{limprob}\lim_{n\rightarrow \infty} \mathbb{P}[\text{broadcast is suppressed}] = \lim_{n\rightarrow \infty} q_n= e^{-1}.\end{equation}
However, it is difficult to extend the analysis to general values of $\alpha$. Therefore, we will resort to a different method, which leverages the particularly nice structure of the matrix $P$.

Instead of considering $\boldsymbol{K^{(n,\alpha)}}$ for general $n$, we will directly consider the case $n\rightarrow \infty$. First, define 
\[ P^{(n,\alpha)}(x,y):=\mathbb{P}\left[K^{(n,\alpha)}_{i+1}/n\leq y\text{ }\bigg\vert\text{ }K^{(n,\alpha)}_i/n=x \right].\]
Then, recalling the structure of the transition matrix $P$, one can deduce for $x,y \in [0,\alpha]$:
\begin{equation}\label{trans}
P^{(\alpha)}(x,y)=\lim_{n\rightarrow \infty} P^{(n,\alpha)}(x,y)=\begin{cases}
\frac{y}{\alpha}, & 0\leq y < \alpha x, \\[1pt]
x, & \alpha x \leq y < \alpha, \\[1pt]
1, & y = \alpha.
\end{cases}
\end{equation}

That is, starting from $x$, the chain moves with probability $x$ to some point in $[0,\alpha x]$ uniformly and with probability $1-x$ it moves to the atom $\alpha$.

We will analyze the Markov chain $\boldsymbol{K^{(\alpha)}}=\{K^{(\alpha)}_i\}_{i=0}^\infty$ with transition function as in \eqref{trans}. For $x<\alpha$, let $\pi_\alpha(x)$ be the steady-state density of  $\boldsymbol{K^{(\alpha)}}$ and let $p_\alpha$ be the steady-state probability of the chain being in the atom $\alpha$.

From \eqref{trans} it is clear that $\pi_\alpha(y)=0$ for $\alpha^2\leq y < \alpha$ and hence for $y<\alpha^2$ we can write
\begin{equation}\label{pi2}
\pi_\alpha(y)=\int_{y/\alpha}^{\alpha^2} \pi_\alpha(x)/\alpha\text{ d}x+p_\alpha / \alpha.
\end{equation}
Moreover, using \eqref{trans} we can write
 \begin{equation}\label{pi1}
p_\alpha=\int_0^{\alpha^2} (1-x)\pi_\alpha(x) \text{ d}x+(1-\alpha)p_\alpha.
\end{equation}
Note, that it follows that $\pi_\alpha(0)=\frac{1}{a}$. Additionally, for $\alpha=1$, Equation \eqref{pi2} is easily solved and we find $\pi_1(x)=e^{-x}$ and, in line with Equation \eqref{limprob}, $p_1=e^{-1}$.

For $\alpha<1$, the solution to \eqref{pi2} can be written recursively by defining $\pi_\alpha(x)$ on distinct intervals as follows
\[
\pi_\alpha(x)=\pi_{\alpha,i}(x)\text{, for }\alpha^{i+1}\leq x < \alpha^{i},
\]
where
\begin{align*}
\pi_{\alpha,1}(x)&=0, \hspace{2cm}\pi_{\alpha,2}(x)=p_\alpha/\alpha, \\
\pi_{\alpha,i}(x)&=\pi_{\alpha,i-1}(\alpha^{i})+\frac{1}{\alpha} \int_{x/\alpha}^{\alpha^{i-1}} \pi_{\alpha,i-1}(y)\text{ d}y, \text{ for } i>2.
\end{align*}

Determining $p_\alpha$ requires a different method. Let $T_\alpha$ be the first return time to the atom $\alpha$. That is
\[
T_\alpha=\min\{i\geq 1:  K^{(\alpha)}_i=\alpha \text{ }\vert\text{ }K^{(\alpha)}_0=\alpha\}.
\]
 We will calculate $\mathbb{E}[T_\alpha]$ and then make use of the fact that $p_\alpha=1/\mathbb{E}[T_\alpha]$. Define
 \[
X^{(\alpha)}_i:= K^{(\alpha)}_i \text{ }\vert\text{ } T_\alpha > i.
\]
Let $f_i(x)$ be the density function of $X^{(\alpha)}_i$. We show that
\begin{equation}\label{pdf}
f_i(x)=\begin{cases}
\frac{i}{\alpha^{i+1}}\left(1-\frac{x}{\alpha^{i+1}}\right)^{i-1}, & 0 \leq x \leq \alpha^{i+1},\\
0,& \text{otherwise.}
\end{cases}
\end{equation}
For $i=1$ we know that $X^{(\alpha)}_1\sim U[0,\alpha^2]$, which is indeed in agreement with \eqref{pdf}. Now suppose \eqref{pdf} is true for some $j$, then
\begingroup
\setlength{\medmuskip}{3mu}
\setlength{\thickmuskip}{4mu}
\begin{align*}
f_{j+1}(y)&=\int_{x=0}^{\alpha^{j+1}}f_j(x)\frac{\text{d}P^{(\alpha)}(x,y)}{\text{d}y}\text{d}x/\mathbb{P}[T_\alpha>j+1\text{\hspace{1pt}}\vert\text{\hspace{1pt}}T_\alpha>j]\\
&=\int_{x=y/\alpha}^{\alpha^{j+1}}f_j(x)\frac{1}{\alpha}\text{d}x/\mathbb{P}[T_\alpha>j+1\text{\hspace{1pt}}\vert\text{\hspace{1pt}}T_\alpha>j].
\end{align*}
\endgroup
Furthermore, as remarked below \eqref{trans}, starting from $x$ the Markov chain moves with probability $x$ to a point in $[0,\alpha x]$, avoiding the return to $\alpha$, and hence
\[
\mathbb{P}[T_\alpha>j+1\text{ }\vert\text{ }T_\alpha>j]=\int_0^{\alpha^{j+1}}f_j(x)x\text{ d}x
=\frac{\alpha^{j+1}}{j+1}.
\]
%=\int_0^{\alpha^{j+1}}\frac{j}{\alpha^{j+1}}\left(1-\frac{x}{\alpha^{j+1}}\right)^{j-1}x\text{ d}x
Combining we find
\[
f_{j+1}(y)=\frac{j+1}{\alpha^{j+2}}\left(1-\frac{x}{\alpha^{j+2}}\right)^{j},
\]
%f_{j+1}(y)=\frac{j+1}{\alpha^{j+1}}.\left(\int_{x/\alpha}^{\alpha^{j+1}}\frac{j}{\alpha^{j+1}}\left(1-\frac{x}{\alpha^{j+1}}\right)^{j-1}\frac{1}{\alpha}\text{ d}x\right)=\frac{j+1}{\alpha^{j+2}}\left(1-\frac{x}{\alpha^{j+2}}\right)^{j},
which proves \eqref{pdf}.

Now noting that for $i>0$ we can write
\[
\mathbb{P}[T_\alpha>i]=\mathbb{P}[T_\alpha>i-1]\mathbb{P}[T_\alpha> i\text{ }\vert\text{ }T_\alpha> i-1],
\]
and using the fact that $\mathbb{P}[T_\alpha> 0]=1$, we deduce
\[\mathbb{P}[T_\alpha> i]=\prod_{j=1}^{i}\frac{\alpha^{j}}{j}=\frac{\alpha^{(i+1)i/2}}{i!}.\]
Finally, we conclude
\[
p_\alpha=\frac{1}{\mathbb{E}[T_\alpha]}=\left(\sum_{i=0}^\infty \mathbb{P}[T_\alpha> i]\right)^{-1}=\left(\sum_{i=0}^\infty\frac{\alpha^{i(i+1)/2}}{i!}\right)^{-1}.\]

%Our last point of interest is the probability that a node other than the central node broadcasts during an arbitrary interval. Using \eqref{pi1} we can deduce
%\begin{equation}
%\lim_{i\rightarrow \infty}\mathbb{E}[K_i^{(\alpha)}]=\int_0^{\alpha^2} x \pi_\alpha(x) \text{ d}x+\alpha p_\alpha=1-p_\alpha.
%\end{equation}
%Hence, on average the central node receives $\frac{n}{\alpha}(1-p_{\alpha})$ transmissions per interval and the asymptotic probability that a node other than the central node broadcasts is 
%$\frac{1}{\alpha}(1-p_{\alpha})$.
\end{document}